# Anisotropic Phase Diagram and Superconducting Fluctuations in SmFeAsO$_{0.85}$F$_{0.15}$

U. Welp[1], C. Chaparro[1], A. E. Koshelev[1], W. K. Kwok[1], A. Rydh[2], N. D. Zhigadlo[3], J. Karpinski[3], S. Weyeneth[4]

[1]Materials Science Division, Argonne National Laboratory, 9700 S. Cass Avenue, Argonne, IL 60439, USA
[2]Department of Physics, Stockholm University, SE-10691 Stockholm, Sweden
[3]Laboratory for Solid State Physics, ETH Zurich, Schafmattstr. 16, CH-8093 Zurich, Switzerland
[4]Physik-Institut der Universität Zürich, Winterthurerstrasse 190, CH-8057 Zürich, Switzerland

We report on the specific heat determination of the anisotropic phase diagram of single crystals of optimally doped SmFeAsO$_{1-x}$F$_x$. In zero-field, the optimally doped compound displays a clear cusp-like anomaly in $C/T$ with $\Delta C/T_c$ = 24 mJ/molK$^2$ at $T_c$ = 49.5 K. In magnetic fields applied along the c-axis, we find pronounced superconducting fluctuations induced broadening and suppression of the specific heat anomaly which can be described using three-dimensional lowest-Landau-level scaling with an upper critical field slope of -3.5 T/K and an anisotropy of $\Gamma$ = 8. The small value of $\Delta C/T_c$ yields a Sommerfeld coefficient $\gamma \sim$ 8 mJ/molK$^2$ indicating that SmFeAsO$_{1-x}$F$_x$ is characterized by a modest density of states and strong coupling.

Following the initial discovery [1] of superconductivity at temperatures up to 26 K in LaFeAsO$_{1-x}$F$_x$, superconductivity has been found in a large number of materials whose common structural motif is the presence of FeAs (or FeSe,Te) planes [2, 3]. Various families of FeAs-superconductors can be distinguished, most notably the (Rare Earth)-1111 materials derived from the original LaFeAsO$_{1-x}$F$_x$, and the 122-family derived from Ba$_{1-x}$K$_x$Fe$_2$As$_2$ [4]. Superconductivity arises upon electron or hole doping or due to the application of pressure from an antiferromagnetic parent compound. The highest values of $T_c$ of ~56 K (resistive onset) were achieved in Sm- and Gd-based 1111-materials [5]. The high values of $T_c$, and the prospect of unconventional s$^{\pm}$-symmetry of the superconducting order parameter, pairing mediated by spin fluctuations and multi-band superconductivity have generated tremendous interest in these new superconductors.

The FeAs-superconductors have distinguishing macroscopic properties such as an enormous upper critical field combined with a small superconducting anisotropy. The upper critical field, $H_{c2}$, its anisotropy and the specific heat anomaly associated with the superconducting transition are fundamental bulk characteristics that shed additional light on the microscopic length scales, the Fermi surface topology and electronic structure of the superconductor.

Here we present the first single crystal specific heat measurements of SmFeAsO$_{0.85}$F$_{0.15}$ to determine the anisotropic phase diagram and the effect of superconducting fluctuations in this material. A clear cusp-like anomaly is observed at the superconducting transition with height of $\Delta C / T_c \approx 24$ mJ/molK$^2$ which is substantially smaller than the prediction based on the scaling $\Delta C / T_c \propto T_c^2$ reported for various Ba-122 based materials [6]. The shape of the zero-field transition and its evolution in applied magnetic fields reveal pronounced superconducting fluctuation effects which can be consistently described in the framework of 3D lowest Landau level (LLL) scaling yielding an upper critical field slope of -3.5 T/K for $H \parallel c$ and a coherence length anisotropy $\Gamma = 8$. The strong superconducting fluctuations are manifested in the very large value of the Ginzburg number $G_i \sim 1.6 \ 10^{-2}$. Entropy conservation and the low value of the specific heat anomaly imply that the Sommerfeld coefficient of the electronic specific heat, $\gamma \sim 8$ mJ/molK$^2$, is lower than previously anticipated, identifying SmFeAsO$_{0.85}$F$_{0.15}$ as a superconductor with modest density of states and strong coupling.

Calorimetric measurements were conducted using a membrane-based steady-state ac-micro-calorimeter [7] with a thermocouple composed of Au-1.7%Co and Cu films deposited onto a 150 nm thick Si$_2$N$_4$-membrane as thermometer. This technique enables high precision measurement of the specific heat of sub-micro gram samples. The absolute accuracy of our specific heat data was checked against gold samples of similar size as our pnictide crystals. SmFeAsO$_{0.85}$F$_{0.15}$ crystals with approximate sizes of 108x95x7 μm$^3$ (sample I) and 130x79x13 μm$^3$ (sample II) were grown in a high-pressure synthesis procedure using NaCl/KCl flux [8]. The samples were mounted onto the thermocouple using Apiezon N grease. An ac-heater current at 23 Hz was adjusted to induce 50 to 200 mK oscillations of the sample temperature. Figure 1 shows the low field magnetization at the superconducting transition of both crystals. The temperature independent magnetization at low temperatures and a transition width of ~ 1.5 K underline the high quality of the crystals.

The inset of Fig. 2a displays the specific heat anomaly near $T_c \sim 49.5$ K of sample I in zero-field. The specific heat is essentially linear in temperature above $T_c$ up to 60 K, the highest temperature measured. We use the linear extrapolation of the normal state specific heat $C_n$ plus a small correction described in detail below as background to analyze the specific heat of SmFeAsO$_{0.85}$F$_{0.15}$ in the temperature range close to $T_c(H)$. At lower temperatures the background specific heat will deviate from linear as the Debye function approaches the characteristic $T^3$-dependence, and the superconducting contribution will be overestimated.

The main panels of Fig. 2 shows the superconducting specific heat $C_s/T$ of sample I in various fields applied along the $c$-axis and $ab$-plane, respectively. Similar data were obtained for sample II. In zero-field a clear almost cusp-like anomaly is observed with a height of ~ 24 mJ/molK$^2$, about twice the value reported on a polycrystalline sample [9] and close to the value of 19 mJ/molK$^2$ obtained on a polycrystalline sample of oxygen deficient F-free SmFeAsO$_{1-x}$ with $T_c$ = 54.6 K [10]. However, our value for $\Delta C/T_c$ is almost an order of magnitude smaller than what would be expected on the basis of the scaling $\Delta C/T_c \propto T_c^2$ that has been reported for various Ba-122 based materials [6]. This indicates that the scaling $\Delta C/T_c = const. \times T_c^2$ is not universal for all FeAs-superconductors *per se*, but that different material families may follow different branches with different values of the constant. The shape of the zero-field anomaly deviates markedly form the conventional mean-field step in $C/T$ at the superconducting transition. Although strong-coupling effects can result in a sharpening of the specific heat anomaly [11], the upward curvature in $C/T$ below $T_c$ the sharp cusp and the long tail above $T_c$ are signatures of strong superconducting fluctuation effects. In magnetic fields applied along the c-axis the peak position, $T_P$, of $C/T$ shifts to lower temperatures and the peak height is strongly suppressed. Concurrently, the onset does not change appreciably, resulting in a strong field-induced broadening of the transition. This field dependence is reminiscent of the behavior seen in cuprate high-$T_c$ superconductors [12] and a further indication of strong fluctuation effects in SmFeAsO$_{0.85}$F$_{0.15}$ as discussed in more detail below. For parallel fields, $H \parallel ab$, this effect is much weaker, indicating strong anisotropy of SmFeAsO$_{0.85}$F$_{0.15}$. As shown in the inset of Fig. 3, the specific heat data in 0.5 T $\parallel$ c virtually superimpose upon those in 4.0 T $\parallel ab$, showing directly that the superconducting anisotropy of SmFeAsO$_{0.85}$F$_{0.15}$ at temperatures near $T_c$ is $\Gamma \sim 8$. This value is in good agreement with previous determinations based on torque magnetometry [13]. For comparison, the companion compound NdFeAsO$_{0.85}$F$_{0.15}$ has an upper critical field anisotropy of $4 - 5$ close to $T_c$ [14, 15].

The measured specific heat, $C$, contains several contributions: $C(T,H) = C_n(T) + C_s(T,H)$, where the normal state background signal $C_n(T) = C_{ph} + \gamma T$ results from phonons and the normal electrons, and the superconducting signal is given as $C_s(T,H) = C_{MF}(T,H) + C_{fl}(T,H)$. Here, $C_{MF}(T,H)$ describes the conventional mean-field step at the superconducting transition, and $C_{fl}(T,H)$ are corrections to the mean-field signal resulting from fluctuation effects. Superconducting fluctuation phenomena may be described using the Ginzburg-Landau free energy functional [16, 17]. Assuming a two-component superconducting order-parameter, the fluctuation contribution to the specific heat (per volume) in zero applied field for a three-dimensional superconductor is given in Gaussian approximation as $C_{fl}(T) = C^+ |t|^{-1/2}$ where $t = (T - T_c)/T_c$, and $C^+ = k_B \Gamma / 8\pi \xi_{ab}^3(0)$ is the amplitude of the fluctuation specific heat for $T > T_c$. For $T < T_c$, the amplitude is $C^- = \sqrt{2}C^+$. $\xi_{ab}(0)$ is the zero-temperature value of the in-plane Ginzburg-Landau coherence length and $\Gamma = \xi_{ab}/\xi_c$ is the coherence length anisotropy. At temperatures very close to $T_c$ the fluctuation contribution may become larger than the mean-field signal, signaling the transition to critical fluctuations and the break-down of the Gaussian approximation. The extent of the critical regime is given by the Ginzburg number $G_i = \left( k_B \mu_0 \Gamma T_c / 4\pi \xi_{ab}^3(0) B_c^2(0) \right)^2 / 2$. Experimentally, the critical regime may be masked by the inhomogeneous broadening of the superconducting transition of a non-ideal sample. In sufficiently strong applied magnetic fields fluctuation effects are enhanced as expressed by the field-dependent Ginzburg number $G_i(H) = \left( H/H_{c2}(0) \right)^{2/3} G_i^{1/3}$. Expressions for the fluctuation specific heat and for other thermodynamic and transport quantities in magnetic fields near $H_{c2}$ can be obtained within the lowest Landau level (LLL) approximation in which the superconducting order parameter is confined to the LLL of the Cooper pairs [18]. This approximation is valid as long as $H > G_i H_{c2}(0)$. These quantities depend on temperature and magnetic field only through scaling variables which for a 3D and 2D superconductor read $(T - T_c(H))/(TH)^{2/3}$ and $(T - T_c(H))/(TH)^{1/2}$, respectively [18].

A challenge in the iThe interpretation of specific heat results has been the fact that the superconducting contribution to the specific heat amounts to only a few percent of the total specific heat, $C_s << C_n$, implying that the normal state background contribution has to be known with very high precision in order to achieve a definitive interpretation of fluctuation effects [19]. Alternatively, the temperature-derivative of the specific heat can highlight the strong temperature variation associated with the superconducting transition over the smooth normal state background. Figure 3 shows the temperature-derivative $dC/dT$ of the total zero-field specific heat. The value of ~0.85 J/molK$^2$ at high temperatures corresponds to the slope of the data in the inset of Fig. 2a. The green lines in Fig. 3 are the fits according to the predictions based on 3D-Gaussian fluctuations, $dC_{fl}/dT = -C^+/2T_c\ |t|^{-3/2}$, yielding the amplitude $C^+ = 71.4$ mJ/molK and $T_{co} = 49.5$ K. This fit describes the data well at temperatures above 50 K. For the fit at $T < T_c$ a linear dependence has been added to account for the temperature dependence of $C_s$ below $T_c$. The integration of the result for $dC_{fl}/dT$ yields, up to a constant, the fluctuation contribution to the specific heat as indicated by the dashed line in Fig. 2a. Its presence indicates that the simple linear extrapolation of the background specific heat has to be corrected by roughly 0.4 % resulting in the data as shown in Figs. 2a and 2b.

The inset of Fig. 2b displays the field dependence of the peak temperature, $T_p$, of the specific heat. Since the onset of the specific heat anomaly is essentially field independent the variation of $T_p$ with field is a measure of the field-dependent width of the transition which comes out to be proportional to $H^{2/3}$. This is the field-dependence expected in Ginzburg-Landau theory for a 3D superconductor suggesting scaling of the in-field specific heat data according to the 3D-LLL-scheme. In analogy to Fig. 3, Fig. 4 shows the data from Fig. 2a in the scaling form of $dC/dT\ (\mu_0 H)^{2/3}\ vs\ (T - T_c(H))/(TH)^{2/3}$ using $\mu_0 dH_{c2}^c/dT = -3.5$T/K. In fields higher than 3 T the data show good scaling, demonstrating that the shape of the in-field specific heat transitions is determined by strong fluctuations in an anisotropic 3D superconductor. In theoretical analysis of the specific heat [18] the scaling properties for the quantity $C_s/C_{MF}$ are obtained. The field and temperature dependences of $C_{MF}$ are not known for Sm-1111; however, experimentally we observe that the coefficient $(\mu_0 H)^{2/3}$ accounts for the field-evolution of the specific heat anomaly very well. A similar relation has been previously found for YBCO [20]. We note though that the scaling property is insensitive to some variability in the upper critical field slope. However, the value of -3.5 T/K deduced from the scaling plot is consistent with the data in Fig. 2a if one chooses the inflection point on the high-temperature side of the transition as $T_c(H)$. Our result is larger than the value obtained from torque magnetometry [13] on a crystal with $T_c \sim 45$ K, which yielded $\mu_0 dH_{c2}^c/dT = -1.9/\eta$ $T/K$ where $\eta$ is a coefficient typically taken to be of order unity. From magneto-transport data on polycrystalline samples [21] a value of $\mu_0 dH_{c2}/dT = -12$ T/K and 2D-scaling of the fluctuation conductivity were deduced. However, the interpretation of such data may be complicated by the mixture of the largely different behaviors for $H \parallel c$ and $H \parallel ab$, respectively, and the obtained critical field slope may in fact be some average of the intrinsic $c$-axis and $ab$-plane data. Magneto-transport on an FIB-patterned SmFeAsO$_{0.70}$F$_{0.25}$ crystal [22] yielded $\mu_0 dH_{c2}^c/dT \sim -3$ T/K if one chooses as criterion the 90%-point of the normal state resistance. For comparison, caloric determinations of the upper critical field slope of NdFeAsO$_{0.85}$F$_{0.15}$ gave values ranging from -0.7 T/K to -2 T/K [13, 14].

Using the standard single-band Ginzburg-Landau relation, the upper critical field slope can be converted into an in-plane coherence length of $\xi_{ab}(0) \approx 1.4$ nm. Combined with the anisotropy coefficient $\Gamma = 8$ this allows for an independent determination of the amplitude of Gaussian fluctuations yielding $C^+ \approx 69$ mJ/molK. This value is in very good − possibly fortuitous - agreement with that obtained from the fit in Fig. 3. In any case, the analysis presented here yields a consistent description of fluctuation effects in SmFeAsO$_{0.85}$F$_{0.15}$ crystals in terms of 3D − GL theory. We estimate a $c$-axis coherence length of $\xi_c(0) = \xi_{ab}(0)/\Gamma \approx 0.14$ nm which is clearly smaller than the repeat distance of the FeAs-layers of $d = 0.85$ nm. Therefore, at low temperatures 2D-superconducting behavior is expected. The crossover temperatures $T_x$ into the 2D-regime in zero-field can be estimated according to $2\xi_c(T_x) \approx d$ [22] yielding $T_x \approx 41.5$ K for $T < T_c$ and $T_x \approx 65.5$ K for $T > T_c$. Thus, the majority of the data shown in Fig. 2 fall into the 3D-regime, consistent with the analysis presented above.

With the help of the general thermodynamic relations $\mu_0 \dfrac{\Delta C_s}{T_c} = \left(\mu_0 \dfrac{dH_c}{dT}\right)^2\bigg|_{T_c} = \dfrac{1}{\beta_A\left(2\kappa^2-1\right)}\left(\mu_0 \dfrac{dH_{c2}}{dT}\right)^2\bigg|_{T_c}$ we can obtain - within Ginzburg-Landau theory - the thermodynamic critical field $B_c(0) = \mu_0 H_c(0) \approx 1.24$ T and the Ginzburg-Landau parameter $\kappa_c \approx 99$. Here, $\beta_A = 1.16$ is the Abrikosov number. Similar to other members of the FeAs-family, SmFeAsO$_{0.85}$F$_{0.15}$ is in the limit of extreme type-II superconductivity. With this value for the thermodynamic critical field a very high Ginzburg number of $G_i \approx 1.6 \times 10^{-2}$ can be deduced which is substantially larger than that reported for other FeAs-superconductors [14, 15, 24] and is a consequence of large anisotropy, high $T_c$ and short coherence lengths.

Entropy conservation yields further constraints on the low-temperature electronic specific heat since the integral of $C_s/T$ taken at temperatures above the zero-crossing of $C_s/T$ equals the integral from zero up to the zero-crossing. The presence of strong fluctuations introduces uncertainty in the evaluation of this integral; however, we believe that the data in Fig. 2a account for the majority of the entropy, ~155 mJ/molK. Although the explicit temperature variation of $C_s$ at low temperatures is not known, with the zero-temperature limit $C_s/T = -\gamma$, a rough estimate based on entropy conservation yields $\gamma \sim 8$ mJ/molK$^2$. Here we consider negligible any residual density of states that might arise due to non-superconducting phase fractions [25] or due to pair-breaking scattering [26]. There is a large variation in reported values of $\gamma$ for SmFeAsO$_{1-x}$F$_x$ ranging from $\gamma \sim 137$ mJ/molK$^2$ [27] to 44 mJ/molK$^2$ [9, 28] and 19 mJ/molK$^2$ [10]. This discrepancy may arise from magnetic contributions to the specific heat associated with the magnetic ordering of the Sm$^{3+}$-ions near 4.5 K. Notwithstanding the uncertainties in our estimate, it appears that such high values of $\gamma$ are inconsistent with the rather small size of the specific heat anomaly at $T_c$. Our results indicate that SmFeAsO$_{1-x}$F$_x$ has a modest value of $\gamma$, i.e., modest density of states $N(E_F)$, which is in contrast to Ba-122 compounds where $\gamma$-values of $\sim$50 mJ/molK$^2$ have been reported [29]. Extensive compilations [3] do show that – on average – the density of states of Ba-122 based compounds is 2 to 3 times larger than that of 1111 and of 11-compunds.

We conclude that SmFeAsO$_{1-x}$F$_y$, and by extension, the other members of the 1111-family, are characterized by a modest density of states and strong coupling which induces high $T_c$. Furthermore, the small value of $\Delta C_s$ promotes a high value of the Ginzburg number, $G_i \sim 1/\Delta C_s^2$, leading to strong fluctuations. The exact value of $G_i$ depends on additional materials parameters such as $\Gamma^2$ and $\xi_{ab}^{-6}$, which in the case of SmFeAsO$_{1-x}$F$_x$ conspire to yield extraordinarily high values of $G_i \sim 1.6 \ 10^{-2}$.

This work was supported by the US Department of Energy – Basic Energy Science – under contract DE-AC02-06CH11357. Work at ETH was supported by the Swiss National Science Foundation NCCR Materials with Novel Electronic Properties (MaNEP) and work at University of Zurich was partially supported by the Swiss National Science Foundation.

Figure captions

Fig. 1: Temperature dependence of the magnetization of samples I and II in a field of 1 mT applied along the *c*-axis measured after zero-field cooling.

Fig. 2: Temperature dependence of the superconducting specific heat of sample I plotted as $C_s/T$ in various magnetic fields applied along the *c*-axis (a) and along the *ab*-planes (b). The dashed line in (a) represents the temperature dependence of the fluctuation contribution in Gaussian approximation. The inset in (a) shows the total specific heat with the solid line indicating the linear background. The inset in (b) shows the peak temperature of $C_s/T$ plotted versus $H^{2/3}$.

Fig. 3: Temperature derivative *dC/dT* of the total zero-field specific heat. The green lines are fits to three-dimensional Gaussian fluctuations. The inset displays the superconducting specific heat $C_s/T$ in units of mJ/molK$^2$ in a field of 0.5 T $\|$ *c* and 4.0 T $\|$ *ab*, respectively, revealing an anisotropy of 8.

Fig. 4: Scaling plot of *dC/dT* $(\mu_0 H)^{2/3}$ versus $(T\text{-}T_c(H))/(TH)^{2/3}$ using an upper critical field slope of -3.5 T/K.

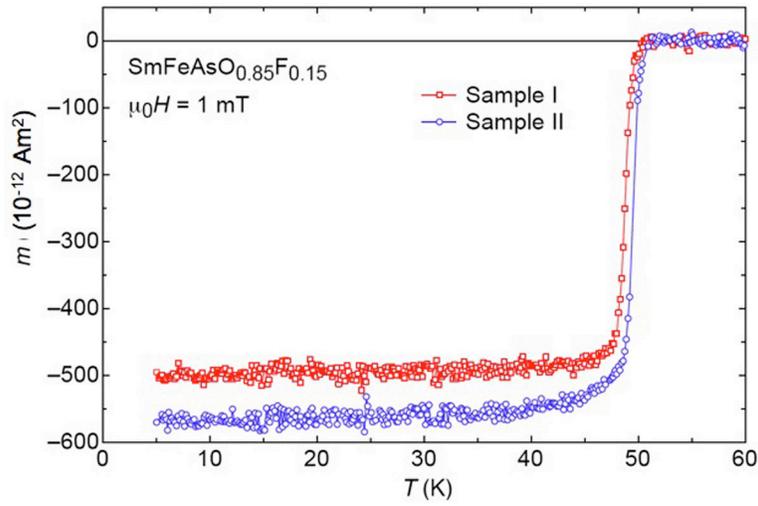

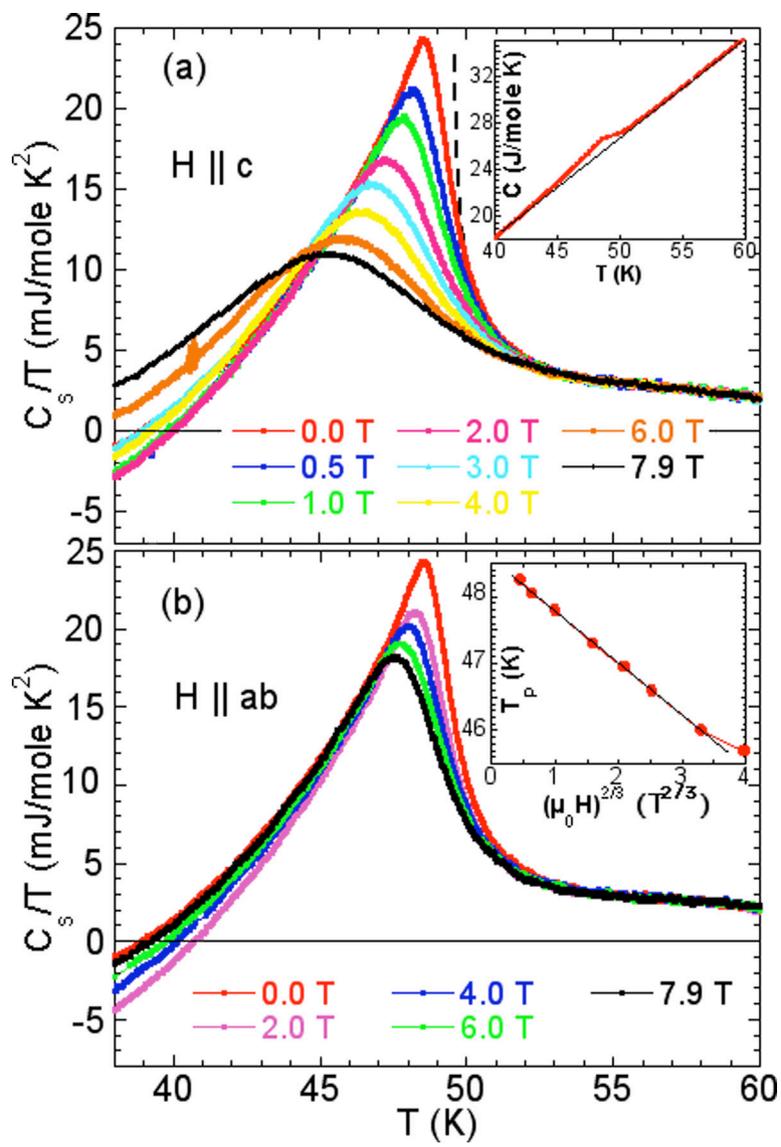

Fig. 2

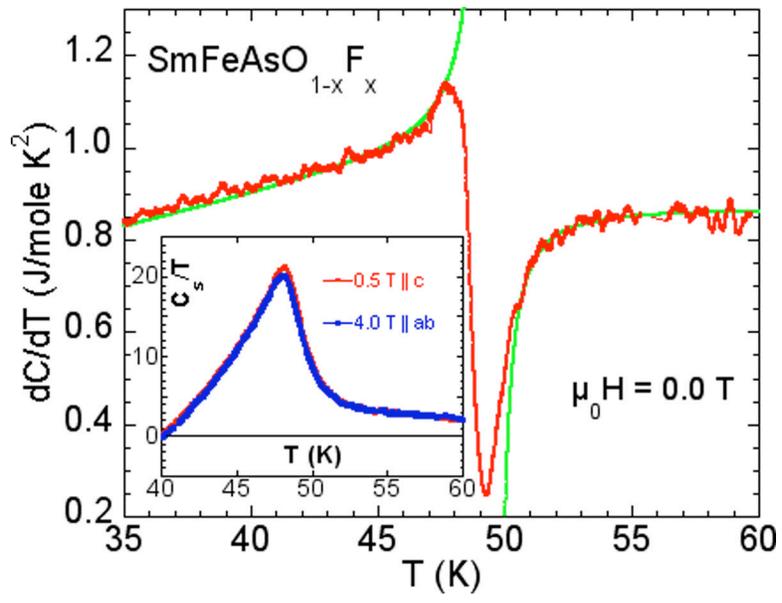

Fig. 3

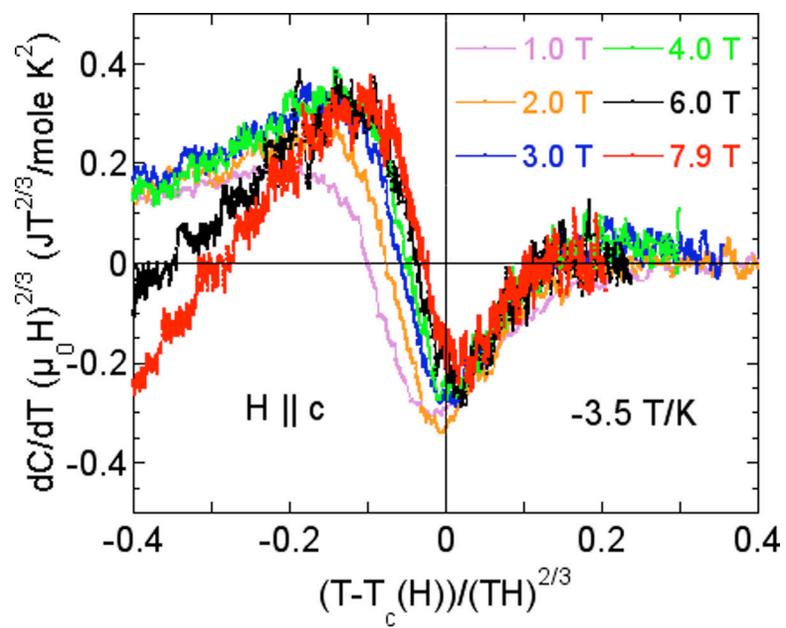

Fig. 4